# Quantum Noise Reduction and Generalized Two-Mode Squeezing in a Cavity Raman Laser

Kai Drühl
Claude Windenberger
Center for Technology Research
Maharishi University of Management
Fairfield, IA 52557
Voice: 515-472-1162
Fax: 515-472-1123
E-mail: kdruhl@mum.edu

August 15, 1998

## Abstract

We study a generalized notion of two-mode squeezing for the Stokes and anti-Stokes fields in a model of a cavity Raman laser, which leads to a significant reduction in decoherence or quantum noise. The model comprises a loss-less cavity with classical pump, unsaturated medium and arbitrary homogeneous broadening and dispersion. Allowing for arbitrary linear combinations of the two modes in the definition of quadrature variables, we find that there always exists a combination of the two output modes which exhibits quadrature squeezing with noise reduction below the vacuum level. The number of noise photons for this combination mode is proportional to the square root of the number of Stokes noise photons.

# 1 Introduction

In this paper, we investigate novel aspects of quantum noise reduction in a model of a linear multi-channel quantum amplifier. Our investigation deals with both state reduction, or decoherence, and squeezing for the amplified channels. In fact, both phenomena turn out to be intimately connected. The model describes stimulated Raman scattering (SRS) in a loss-less cavity with a single Stokes mode, a single anti-Stokes mode and a classical pump. We assume the Raman medium to be not saturated, but allow for arbitrary homogeneous broadening and dispersion. The Raman amplification process gives rise to both two-mode squeezing and decoherence for the two optical field modes. We find that quantum noise resulting from these processes can be largely cancelled by appropriate interference arrangements, while maintaining the full amount of amplification.

The first arrangement is an 'anti-squeezing' operation, by which two uncorrelated normal modes are obtained. This operation can be implemented, for example, by a linear parametric amplifier. If the initial state of fields, medium and reservoir is the vacuum, the state for one of the normal modes is the vacuum, while for the other it is a canonical ensemble of number states. The average number of "noise" photons for the latter mode is equal to the number of phonons in the medium and reservoir states.

The second arrangement involves linear mixing of Stokes and anti-Stokes without additional squeezing. The state of the resulting mixed mode is a canonical ensemble of squeezed number states. For a suitable, optimal choice of mode coefficients, the variance of one of the quadrature phases [1] of the mode is below the vacuum level, in spite of the additional quantum noise resulting from decoherence of the mixed state. This result holds for all values of dispersion, line width and coupling constants. The variance of the conjugate quadrature is comparable to the variances of the Stokes quadratures. For large amplification, the number of squeezed photons in the mixed mode is proportional to the square root of the number of Stokes photons. This represents a significant reduction in the amount of decoherence for the mixed mode, as compared to the original modes.

The use of general mode coefficients amounts to a generalization of the notion of two-mode squeezing. For conventional two-mode squeezing, both mode coefficients are chosen to be of equal magnitude. (This is the definition reported by Loudon and Knight [1]. Caves and Schumaker rescale the coefficients by frequency dependent factors, to correct for the frequency dependent relationship between photon and field operators [2].) In this case, the variances of both quadratures and the number of photons of the mixed mode grow in proportion to the number of Stokes photons, indicating a much higher level of quantum noise than for the optimal choice of mode coefficients.

The reduction of quantum noise through squeezing has been studied extensively [1,3]. A squeezed state of a harmonic oscillator is a state for which the variance in one of the conjugate variables is reduced at the expense of an increase in the variance of the other variable. (We will not discuss the more general notions of higher order squeezing which have been defined in terms of the variance of polynomials of conjugate variables [4,5]. In the context of quantum optics it has become customary to speak of conjugate variables as quadrature phases [1]).

For example, the variance in the electric field of an optical mode may be reduced below its vacuum level at the expense of an increased variance in the magnetic field. This situation is desirable for high sensitivity interferometer experiments [1,6] and optical communications [1,7]. Such squeezed states of the electromagnetic field can be generated by non-linear optical processes, such as degenerate parametric amplification or four-wave mixing, in which photons are created in pairs. If the process is non-degenerate, i.e. if two photons at different frequencies are generated, squeezing is not found for each of the individual modes, but rather for linear combinations of the two. This is known as two-mode squeezing, and its experimental detection requires an additional coherent source at the sum- or difference frequency [1]. Two-mode squeezing is conventionally defined in terms of linear combinations with equal mode coefficients [1,2]. We will show in this paper that, for situations where decoherence occurs, this is not an optimal definition, and that it is useful to adjust the mode coefficients to compensate for different noise levels in the two modes.

The generation of squeezed light involves an amplification process, and this is frequently associated with state reduction or decoherence. State reduction (the transformation of a pure state into a statistical mixture) occurs due to the coupling of the observed modes to a number of unobserved or unobservable modes. The process of state reduction in linear quantum amplifiers has been studied for various types of models, and the following references represent just a small selection of papers which are of interest here [8,9,10]. Such models are of great interest for the description of dissipation in quantum systems [9,10], and for the analysis of state reduction in quantum measurement theory [8,10]. Most studies refer to harmonic oscillator models involving a small number of observable variables, and a large or infinite number of unobserved 'reservoir' oscillators. The time evolution of the amplifier can be described either by master equations for the observable degrees of freedom [10], or by exact solutions for the linear equations of motion of the complete system [8,9,10].

The exact solutions reveal that, for these models, the time evolution is a multi-mode squeezing transformation coupling the observed variables and the reservoir. The amount of squeezing in the observable variables depends on the type of model studied. If the coupling to the reservoir involves only one of the conjugate variables, an initial rapid buildup of decoherence and squeezing for that variable occurs on a very short time-scale, of the order of the reservoir coherence time [9,10]. In addition, decoherence for both conjugate variables increases on a much longer time scale, given by the dissipation rate of the reservoir. In quantum optics it is customary to employ the 'rotating wave approximation' (RWA), in which the coupling involves both conjugate variables of the observable system in a symmetric fashion. In this approximation, single-mode squeezing is absent, and decoherence occurs on the longer, dissipative time scale [9,10]. The cavity SRS amplifier studied here is in this category.

So far, studies of quantum amplifiers in the RWA have emphasized either the process of state reduction, or the generation of squeezed states. However, both aspects are intimately connected. The authors found in [11] that, for the simplest case of a single mode amplifier coupled to an infinite reservoir, the time evolution results in two-mode squeezing for the amplifier mode and a single, time-dependent reservoir mode. If only the amplifier mode is observed, no squeezing is found, and the density matrix contains a large number of pure components, proportional to the amplification factor. If, on the other hand, both the amplifier mode and the large-time limit of the special reservoir mode are observed, the density matrix contains only a small number of squeezed pure components. Similar results were found for a model of a single mode laser, comprising a single radiation mode, an inverted medium and a reservoir. To the best of our knowledge, these connections between the choice of observed mode combinations, the degree of state reduction and the degree of squeezing have not been systematically explored so far.

The linear cavity Raman amplifier studied here is of particular interest,

both because of the type of interactions found in this system, and as an object for experimental study. The medium is assumed to contain a large number of unsaturated Raman active molecules or atoms, and is modelled by a single oscillator mode. The coupling between Stokes and medium oscillator is of squeezing type. However, since the medium is not directly observable, two-mode variables constructed from these two modes have no experimental significance. On the other hand, the coupling between anti-Stokes and medium conserves photon-phonon number, and the anti-Stokes mode can therefore be considered as a probe of the medium. In fact, the squeezing transformation occurring between Stokes and medium is reflected in corresponding correlations between Stokes and anti-Stokes. While the coupling between anti-Stokes and medium is in most cases stronger than the Stokes coupling, the influence of the anti-Stokes on the medium is reduced by dispersion, which leads to a frequency detuning of the anti-Stokes. It is therefore of interest to study the behavior of the CRA over the whole range of relevant parameters, including dispersion, homogeneous line width and gain. Experimentally, all of these parameters can be controlled independently, by varying the pressure of the gaseous Raman medium, the type of medium and the pump intensity.

Cavity SRS models with Stokes and anti-Stokes modes have been studied by several authors. Non-linear, micro-maser models of SRS in a cavity with a single three-level atom and without dispersion were studied by Law et al. [12] and Puri et al. [13]. The latter found noise reduction below the vacuum level with the conventional notion of two-mode squeezing for certain values of the number of passes and the passing time of the atom through the cavity. Linear oscillator models of the medium and reservoir had been studied earlier by Walls [14] and Perina [15] for arbitrary dispersion and for the limiting case of large line width, where the medium oscillators are eliminated adiabatically. These studies were mainly concerned with photon statistics, and give no explicit discussion of squeezing. The first detailed study of squeezing in this type of models was done by Chizhov et al., who studied quadrature and higher order squeezing for the case of zero dispersion in a cavity [16,17]. A free-propagation amplifier was studied by Yeong et. al. [18]. In the limit of large line width [16], quadrature squeezing was reported to be absent, while such squeezing was found for the case of small line width in [17] and [18]. These studies discuss conventional two-mode quadrature squeezing in terms of the variance of one of the two-mode quadratures.

Our present work studies the linear cavity SRS model for arbitrary values of line width and dispersion. (The case of zero dispersion is rather special, in that Stokes, anti-Stokes and the medium undergo damped oscillations. For non-zero dispersion, exponential amplification occurs.) Two-mode quadrature squeezing is defined in terms of the ratio of the variances of two conjugate two-mode quadratures, with arbitrary mode coefficients. We give explicit expressions for the Wigner characteristic function of the entangled state of Stokes and anti-Stokes modes, and determine those values of the mode coefficients, for which maximal squeezing and minimal state-reduction occur.

In section 2.1, we give the model equations for the amplitude operators, derive the general form of solutions, and calculate the variance of a general Hermitean combination of the two optical modes for the case where the initial state at time t=0 is the vacuum. As is well known, this uniquely determines the Wigner characteristic function, and thereby the Schrödinger state at later times $t > 0$. In section 2.2, we show that the reduced state of Stokes and anti-Stokes is a product of the vacuum and a canonical ensemble of number states for two uncorrelated normal modes. These normal modes are obtained from the original optical modes by an 'anti-squeezing' transformation, and we show that this transformation can be realized by a linear parametric amplifier. In section 2.3, we study 'non-squeezed' linear combinations of the two optical modes, and generalize the corresponding notion of two-mode squeezing by admitting arbitrary mode coefficients. We show that, at any given time t, it is possible to choose optimal mode coefficients in such a way, that one of the quadratures has minimal variance below the vacuum level at that time. This choice also leads to minimal decoherence. In situations where the system shows exponential amplification, the number of photons in the optimal mode is proportional to the square root of the number of Stokes photons, indicating a much higher level of quantum coherence than for the Stokes mode. In section 2.4, we consider amplification of an initial coherent state of the Stokes mode. The coherent state is maximally amplified if it is at 90 degrees relative to the minimal two-mode quadrature, and gives the same output as the vacuum if it is parallel to that quadrature. In the former case, amplitude fluctuations are enhanced, while phase fluctuations are below the vacuum level.

In section 3.1 we give the general solution to the linear equations of motion for optical modes and medium in terms of the eigenvalues of the 3x3 matrix of coefficients. We discuss the special cases of zero and large line width in sections 3.2 and 3.3, and the intermediate case of finite line width in section 3.4. In particular, we consider the influence of dispersion on the long-time behavior of the system. For zero line width, there exists a critical value of dispersion, below which the system shows periodic or quasi-periodic behavior, and above which exponential amplification occurs. For non-zero line width, the system always shows exponential amplification at large times, except for zero dispersion, where the behavior is that of damped oscillations.

In section 4, we summarize our findings and discuss some possible implications for a deeper understanding of state reduction in more general systems.

# 2 Model equations, state reduction and squeezing

## 2.1 Model equations and characteristic functions

Our model of a cavity Raman laser contains the following variables: a classical pump at frequency $\omega_0$, a Raman-active medium with transition frequency $\omega_3$, a reservoir of oscillators with frequencies $\omega_r, r = 4...N$, modelling homogeneous broadening of the Raman transition, a Stokes mode at frequency $\omega_1 = \omega_0 - \omega_3$, and an anti-Stokes mode at frequency $\omega_2 = \omega_0 + \omega_3 + \Delta$. The parameter $\Delta$ is a frequency detuning of the anti-Stokes mode which results from dispersion in the Raman-active medium. Figure 1 shows a level diagram of the medium, including the Raman transition, two transient intermediate states and the three optical frequencies $\omega_0$, $\omega_1$, and $\omega_2$.

The medium is assumed to be only weakly excited, with most of its population in the ground state. It can therefore be modelled by a single harmonic oscillator. We write the Heisenberg operators $\tilde{b}_n(t)$, $n = 1...N$ for all variables involved as products of time-dependent phase factors and amplitude operators $b_n(t)$:

$$\begin{aligned}
\tilde{b}_1(t) &= e^{-i(\omega_0-\omega_3)t} b_1(t) , \\
\tilde{b}_2(t) &= e^{-i(\omega_0+\omega_3)t} b_2(t) , \\
\tilde{b}_k(t) &= e^{-i\omega_k t} b_k(t) \qquad (k = 3, ..., N) .
\end{aligned}$$

The Heisenberg equations of motion for the amplitude operators in the

Markov approximation are [16,17]:

$$\dot{b}_1^\dagger = \kappa_1 b_3 , \tag{1}$$

$$\dot{b}_2 = -i\Delta b_2 - \kappa_2 b_3 , \tag{2}$$

$$\dot{b}_3 = \kappa_1 b_1^\dagger + \kappa_2 b_2 - \Gamma b_3 + F, \qquad F = \sum_{s=4}^N g_s e^{-i\Delta_s t} b_s(0) , \tag{3}$$

$$\dot{b}_r = -g_r e^{i\Delta_r t} b_3 \qquad (r = 4...N) , \tag{4}$$

where $\kappa_1$, $\kappa_2$ and $g_r$ are the coupling coefficients for the medium with the Stokes, anti-Stokes and reservoir variables. The frequency detunings are $\Delta = \omega_2 - \omega_3 - \omega_0$ and $\Delta_r = \omega_r - \omega_3$. The solutions of these equations take the form:

$$b_1^\dagger(t) = T_{11}(t) a_1^\dagger + \sum_{k=2}^N T_{1k}(t) a_k ,$$

$$b_i(t) = T_{i1}(t) a_1^\dagger + \sum_{k=2}^N T_{ik}(t) a_k \qquad (i = 2...N) , \tag{5}$$

where $a_n = b_n(0)$, $n = 1...N$, and the coefficients $T_{nm}(t)$ are c-number solutions of equations (1) through (4) which satisfy:

$$T_{nm}(0) = 1; \qquad n = m$$
$$T_{nm}(0) = 0; \qquad n \neq m$$

Because equation (3) for $b_3$ involves only the initial values $b_r(0)$, $r = 4...N$, of the reservoir variables, the coefficients $T_{ik}(t)$ for $i, k = 1, 2, 3$ can be calculated from equations (1), (2) and a modified version of equation (3), for which $F = 0$:

$$\dot{b}_3 = \kappa_1 b_1^\dagger + \kappa_2 b_2 - \Gamma b_3 . \tag{6}$$

In the following, we will study the coherence properties of the Stokes and anti-Stokes modes in the state $|\psi(t)>$ at time $t$. The state is uniquely characterized by its Wigner characteristic function $\chi$, depending on $t$ and arbitrary complex coefficients $\alpha_n$:

$$\chi(t, \alpha_n, \alpha_n^*) = <\psi(t)|e^{i[\phi(0)+\phi^\dagger(0)]}|\psi(t)>$$
$$= <\psi(0)|e^{i[\phi(t)+\phi^\dagger(t)]}|\psi(0)> \tag{7}$$

where the operator $\phi(t)$ is defined by:

$$\begin{aligned}\phi(t) &= \alpha_1^* b_1^\dagger(t) + \sum_{j=2}^{N} \alpha_j b_j(t) \\ &= \beta_1^*(t) a_1^\dagger + \sum_{k=2}^{N} \beta_k(t) a_k \end{aligned} \quad (8)$$

From equation (5), the time-dependent coefficients $\beta_n$ are given by:

$$\begin{aligned}\beta_1^*(t) &= \alpha_1^* T_{11}(t) + \sum_{i=2}^{N} \alpha_i T_{i1}(t) \\ \beta_k(t) &= \alpha_1^* T_{1k}(t) + \sum_{i=2}^{N} \alpha_i T_{ik}(t) \qquad (k=2...N)\end{aligned} \quad (9)$$

From the canonical commutation relations for the amplitude operators, one finds:

$$[\phi(t), \phi^\dagger(t)] = -|\beta_1|^2 + \sum_{k=2}^{N} |\beta_k|^2 = -|\alpha_1|^2 + \sum_{j=2}^{N} |\alpha_j|^2 \quad (10)$$

If the initial state $|\psi(0)>$ is a product of vacuum states for all amplitude operators $a_n$, the characteristic function $\chi$ takes the form

$$\chi(t, \alpha_n, \alpha_n^*) = e^{-\frac{1}{2}\varphi^2(t)}$$

where

$$\begin{aligned}\varphi^2(t) &= <\psi(0)|(\phi(t)+\phi^\dagger(t))^2|\psi(0)> \\ &= \sum_{n=1}^{N} |\beta_n|^2 = 2|\beta_1|^2 - |\alpha_1|^2 + \sum_{i=2}^{N} |\alpha_i|^2\end{aligned} \quad (11)$$

Here we have used equation (10) to eliminate the dependence on all coefficients $\beta_i$ with $i > 1$. If only the Stokes and anti-Stokes modes are observed, we set $\alpha_i = 0$ for $i > 2$, and obtain

$$\begin{aligned}\varphi^2(t) &= 2|\beta_1|^2 - |\alpha_1|^2 + |\alpha_2|^2 \\ \beta_1^*(t) &= \alpha_1^* T_{11}(t) + \alpha_2 T_{21}(t)\end{aligned} \quad (12)$$

Equations (11) and (12) for the characteristic function of the two optical modes are the main results of this subsection. They show that the reduced state of these modes is completely determined by the two complex coefficients $T_{11}$ and $T_{21}$. For observation of the Stokes mode alone, one sets $\alpha_2 = 0$ and finds

$$\varphi^2(t) = (2|T_{11}|^2 - 1)|\alpha_1|^2 = (2n_1 + 1)|\alpha_1|^2 \qquad (13)$$

For observation of the anti-Stokes mode alone, one sets $\alpha_1 = 0$ and finds

$$\varphi^2(t) = (2|T_{21}|^2 + 1)|\alpha_2|^2 = (2n_2 + 1)|\alpha_2|^2 \qquad (14)$$

Here $n_i$ is the average photon number:

$$n_i = <\psi(0)|\, b_i^\dagger b_i \,|\psi(0)> = \begin{cases} |T_{11}|^2 - 1 & \text{for } i = 1 \\ |T_{i1}|^2 & \text{for } i > 1 \end{cases}$$

The characteristic function (11) of each of these modes is a canonical ensemble of number states, which is uniquely characterized by $n_i$ (see Appendix A.1). From equation (10), the following relationship is derived for the coefficients $T_{i1}$

$$|T_{11}|^2 - |T_{21}|^2 = \sum_{k=3}^{N} |T_{k1}(t)|^2 \equiv R^2 \geq 1 \qquad (15)$$

In terms of photon numbers, this is equivalent to

$$n_1 - n_2 = \sum_{k=3}^{N} n_k = R^2 - 1 \geq 0 \qquad (16)$$

showing that the difference between Stokes and anti-Stokes photon numbers is equal to the total number of medium and reservoir phonons.

## 2.2 "Anti-squeezing", Normal Modes and Decoherence

For $t > 0$, the Stokes and anti-Stokes modes are correlated. Equating the coefficients of $\alpha_1 \alpha_2$ in equations (11) and (12), we find:

$$<\psi(0)|\, b_1 b_2 \,|\psi(0)> = T_{11}^* T_{21}. \qquad (17)$$

We now seek to write the two modes $b_i$ as linear combinations of uncorrelated normal modes $c_i$:

$$b_1^\dagger = Uc_1^\dagger + V^*c_2 \qquad (18)$$
$$b_2 = Vc_2^\dagger + U^*c_1 \quad ; \quad |U|^2 - |V|^2 = 1$$

In terms of the normal modes $c_i$, the mode operator $\phi$ of equation (8) takes the form:

$$\phi(t) = \alpha_1^* b_1^\dagger + \alpha_2 b_2 = \gamma_1^* c_1^\dagger + \gamma_2 c_2 \qquad (19)$$

with complex coefficients $\gamma_i$ given by

$$\gamma_1^* = \alpha_1^* U + \alpha_2 V ,$$
$$\gamma_2 = \alpha_1^* V^* + \alpha_2 U^* , \qquad (20)$$

and

$$-|\alpha_1|^2 + |\alpha_2|^2 = -|\gamma_1|^2 + |\gamma_2|^2$$

Equation (12) for the variance $\varphi^2$ suggests to choose

$$U = R^{-1} T_{11} \quad , \quad V = R^{-1} T_{21} . \qquad (21)$$

With this choice, $\varphi^2$ is diagonalized:

$$\varphi^2 = (2R^2 - 1)\gamma_1^2 + \gamma_2^2 \qquad (22)$$

and the characteristic function $\chi(\gamma_1, \gamma_2)$ of the two normal modes factorizes:

$$\chi(\gamma_1, \gamma_2) = e^{-\frac{1}{2}(2R-1)|\gamma_1|^2} e^{-\frac{1}{2}|\gamma_2|^2} . \qquad (23)$$

This shows that the reduced state for the two normal modes is the product of a canonical ensemble of number states for $c_1$ (see appendix A.1) and the vacuum state for $c_2$. In particular, the normal modes are uncorrelated and not squeezed. The number $n_1'$ of photons in mode $c_1$ is equal to

$$n_1' = R^2 - 1 = n_1 - n_2 . \qquad (24)$$

For the special situation where $R^2 = 1$, the coefficients $T_{i1}$ for $i > 2$ vanish. The normal modes $c_i$ are then just the Schrödinger operators $b_i(0) = a_i$, and equation (18) is the original time-evolution, equation (5). As we shall see below, this situation occurs for $\Gamma=\Delta=0$ periodically at certain times $t > 0$.

It is remarkable that, even for $R^2 > 1$, one of the normal modes is always found in the vacuum state.

A physical interpretation of the normal modes $c_i$ can be given in terms of an "anti-squeezing" experiment, performed on the output of the cavity Raman laser. Inverting equations (18), we find

$$\begin{aligned} c_1^\dagger &= U^* b_1^\dagger - V^* b_2 \\ c_2 &= -V b_1^\dagger + U b_2 \end{aligned} \quad (25)$$

This is a squeezing operation performed on the modes $b_1$ and $b_2$, which can be implemented, for example, by a non-degenerate linear parametric amplifier (LPA) [3]. If the Stokes and anti-Stokes modes are used as idler and signal in a phase-matched linear amplifier, the equations for propagation in the z-direction are [3]:

$$\begin{aligned} \frac{d}{dz} c_1^\dagger &= g e^{-i\eta} c_2 ; \qquad c_1^\dagger(0) = b_1^\dagger \\ \frac{d}{dz} c_2 &= g e^{i\eta} c_1^\dagger ; \qquad c_2(0) = b_2 \end{aligned} \quad (26)$$

where $\eta$ is the pump phase, and $g$ is the product of pump amplitude and gain coefficient. The solutions are:

$$\begin{aligned} c_1^\dagger(z) &= \cosh gz \, b_1^\dagger + e^{-i\eta} \sinh gz \, b_2 \\ c_2(z) &= e^{i\eta} \sinh gz \, b_1^\dagger + \cosh gz \, b_2 \end{aligned} \quad (27)$$

Equation (25) is obtained, apart from a common phase factor of $c_1$ and $c_2$, for

$$\cosh gz = |U| ; \qquad \eta = \arg(V) - \arg(U) \quad (28)$$

Equations (19) and (23) show that, at any given time $t$, this operation will disentangle the state of Stokes and anti-Stokes modes, leading to zero correlation between the new modes $c_1$ and $c_2$ and to zero photon count for mode $c_2$. We illustrate the sequence of squeezing and "anti-squeezing" operations performed by the cavity Raman laser (CRL) and the LPA in figure 2.

By adjusting the pump amplitude and phase of the LPA, arbitrary "anti-squeezing" transformations can be performed on the optical fields. For example, if the pump phase $\eta$ is changed by an amount $2\epsilon$ from the value given in (28), we find for the number of photons in the exit mode $c_2(z)$:

$$\eta = \arg(V) - \arg(U) + 2\epsilon \tag{29}$$

and

$$<\psi(0)|\, c_2^\dagger(z) c_2(z) \,|\psi(0)> \;=\; 2\frac{(n_1+1)n_2}{1+n_1-n_2}\sin^2\epsilon \;>\; 2n_2\sin^2\epsilon. \tag{30}$$

For $n_2 \gg 1$, a single photon is counted for $\epsilon = (2n_2)^{-\frac{1}{2}}$, and a very stringent condition on $\eta$ is obtained. Similar restrictions are found for the pump amplitude $g$.

## 2.3 Generalized two-mode Squeezing for Stokes and anti-Stokes

Two-mode squeezing for two oscillators $b_1$ and $b_2$ is conventionally defined in terms of the variance of the Hermitian components [1]:

$$
\begin{aligned}
X_Q &= \frac{1}{2}(b_1 + b_1^\dagger + b_2 + b_2^\dagger) \\
X_P &= \frac{-i}{2}(b_1 - b_1^\dagger + b_2 - b_2^\dagger)
\end{aligned}
\tag{31}
$$

These operators are just the quadratures of a linear combination $b$ of $b_1$ and $b_2$:

$$b = \frac{1}{\sqrt{2}}b_1 + \frac{1}{\sqrt{2}}b_2 \tag{32}$$

$$X_Q = \frac{1}{\sqrt{2}}(b + b^\dagger) \;;\qquad X_P = \frac{-i}{\sqrt{2}}(b - b^\dagger)\,. \tag{33}$$

A state is two-mode squeezed for $b_1$ and $b_2$ if it is squeezed for $b$. The conventional definition of squeezing is that the variance of one of the Hermitian components is below the vacuum value $\frac{1}{2}$:

$$<(\Delta X_F)^2> \;<\; \frac{1}{2} \qquad \text{for } F = P \text{ or Q}\,. \tag{34}$$

Because of the uncertainty relations, this implies that

$$<(\Delta X_P)^2> \neq <(\Delta X_Q)^2> . \tag{35}$$

In this paper, we use the term "squeezed state" for states satisfying the weaker condition (35). If the state is a minimal uncertainty state, e.g. the vacuum state, one has

$$<(\Delta X_P)^2><(\Delta X_Q)^2> = \frac{1}{4} \tag{36}$$

and equation (35) implies (34). For states satisfying equation (35) but not (36), however, the variance of both components may exceed the vacuum value.

To generalize the notion of two-mode squeezing, we first admit arbitrary coefficients $z_i$ in the definition of the linear combination $b$

$$b = z_1 b_1 + z_2 b_2 \quad ; \quad |z_1|^2 + |z_2|^2 = 1 . \tag{37}$$

Second, we allow arbitrary phase factors in the definition of quadratures. We thus consider a general Hermitian linear combination of $b$ and $b^\dagger$ as in the definition (7) and (8) of the characteristic function

$$\phi_1 = \alpha^* b^\dagger + \alpha b \tag{38}$$

For $\alpha = \frac{1}{\sqrt{2}}$ or $\alpha = \frac{-i}{\sqrt{2}}$, one has $\phi_1 = X_Q$ or $\phi_1 = X_P$. To determine the degree of squeezing of a given state, one needs to find those values of $\alpha$ with $|\alpha| = \frac{1}{\sqrt{2}}$, for which the variance of $\phi_1$ assumes maximal and minimal values $\varphi_+^2$ and $\varphi_-^2$. We now apply this procedure to the Stokes and anti-Stokes modes defined above.

From equations (38) and (8) we find that the characteristic function of the mode $b$ in equation (37) is obtained from the general expressions (11) and (12) by setting $\alpha_1 = \alpha z_1$ and $\alpha_2 = \alpha z_2$. Thus

$$\beta_1^* = \alpha^* z_1^* T_{11} + \alpha z_2 T_{21} . \tag{39}$$

This gives for the variance $\varphi_1^2$ of $\phi_1$ (see equation 12):

$$\varphi_1^2 = 2|\alpha^* z_1^* T_{11} + \alpha z_2 T_{21}|^2 + |\alpha|^2(|z_2|^2 - |z_1|^2) . \tag{40}$$

To quantify the amounts of state reduction and squeezing, we rewrite this in the form (see Appendix A.1)

$$\varphi_1^2 = Q^2|\alpha^*U' + \alpha V'|^2; \qquad |U'| = \cosh s \, , \, |V'| = \sinh s. \qquad (41)$$

This corresponds to a canonical ensemble of squeezed number states with average photon number $n = \frac{1}{2}(Q^2 - 1)$ and squeezing parameter $s$. For $|\alpha| = \frac{1}{\sqrt{2}}$, we find the maximal and minimal variances of the corresponding quadratures from equations (40) and (41):

$$\varphi_\pm^2 = (r_1R_1 \pm r_2R_2)^2 + \frac{1}{2}(r_2^2 - r_1^2) = \frac{1}{2}Q^2 e^{\pm 2s} \, , \qquad (42)$$

$$\text{where} \qquad r_i = |z_i| \qquad \text{and} \qquad R_i = |T_{i1}| \, . \qquad (43)$$

This gives for the coefficients $Q^2$ and $e^{-2s}$

$$Q^2 = 2\sqrt{\varphi_-^2 \cdot \varphi_+^2} \, , \qquad e^{-2s} = \sqrt{\varphi_-^2 \cdot \varphi_+^{-2}} \qquad (44)$$

We now consider a situation where

$$R_1, R_2 \to \infty \, , \qquad \frac{R_2}{R_1} \to q < 1 \text{ as } t \to \infty \, . \qquad (45)$$

and give the results for various choices of mode coefficients $z_i$.

Case 1: The conventional choice of mode coefficients is $|z_i| = \frac{1}{\sqrt{2}}$ and gives:

$$\varphi_\pm^2 = \frac{1}{2}(R_1 \pm R_2)^2 \to \frac{1}{2}R_1^2(1 \pm q) \, ,$$

$$Q^2 = R_1^2 - R_2^2 \to R_1^2(1 - q^2) \, , \qquad e^{-2s} \to \frac{1-q}{1+q} \qquad (46)$$

In this case, the quantum noise in both quadratures increases as $R_1^2$, and goes above the vacuum level. The average squeezed photon number increases in proportion to $R_1^2$, while the squeezing coefficient goes to a finite limit.

Case 2: If the mode coefficients $z_i$ are adjusted at any given time to make the first term in the expression for $\varphi_-^2$ in equation (42) vanish, we obtain:

$$|z_1^*T_{11}| = |z_2 T_{21}| \, , \qquad (47)$$

$$\varphi_-^2 = \frac{1}{2}(r_2^2 - r_1^2) \to \frac{1}{2}(\frac{1-q^2}{1+q^2}) < \frac{1}{2} \quad , \quad \varphi_+^2 \to 4r_1^2 R_1^2 = 4R_1^2 \frac{q^2}{1+q^2} \, ,$$

$$Q^2 \to 2\sqrt{2} R_1 \frac{q}{1+q^2}\sqrt{1-q^2} \quad , \quad e^{-2s} \to \frac{1}{2\sqrt{2}} R_1^{-1} \frac{1}{q}\sqrt{1-q^2} \, .$$

In this case, the quantum noise in the quadrature corresponding to $\varphi_-^2$ stays below the vacuum limit. The average squeezed photon number increases in proportion to $R_1$, while the squeezing coefficient decreases in proportion to $R_1^{-1}$.

Case 3: It is possible to choose the mode coefficients $z_i$ at any given time such as to minimize the variance $\varphi_-^2$. This minimal value is given by:

$$\varphi_{min}^2 = \frac{1}{2}\left(R_1^2 + R_2^2 - \sqrt{(R_1^2 - R_2^2 - 1)^2 + 4R_1^2 R_2^2}\right) < \frac{1}{2} \qquad (48)$$

In the limit of large times where equation (45) holds, this gives the same results as for the choice of coefficients in equation (47) above.

Case 4: It turns out that, in many situations, noise reduction below the vacuum level can be achieved with constant mode coefficients and quadratures. For the choice $\alpha = \frac{-i}{\sqrt{2}}$, we get for the variance in this case

$$\varphi_{fix}^2 = |z_1^* T_{11} - z_2 T_{21}|^2 + \frac{1}{2}(|z_2|^2 - |z_1|^2) \, . \qquad (49)$$

## 2.4 Amplification of coherent states

In this section, we calculate the final state for the linear combination mode, as discussed in section (2.3), for the case where the initial state is a coherent state for the Stokes field, and the vacuum state for all other variables.

$$a_1|\psi(0)> = \nu_1|\psi(0)> \quad , \quad a_i|\psi(0)> = 0 \text{ for } i > 1 \qquad (50)$$

In this case, the characteristic function takes the form

$$\chi(t, \alpha, \alpha^*) = e^{i(\beta_1 \nu_1 + \beta_1^* \nu_1^*) - \frac{1}{2}\varphi_1^2} \, . \qquad (51)$$

If we take the near-optimal choice of case 2 for the mode coefficients $z_i$, for which

$$|z_1^* T_{11}| = |z_2 T_{21}| = r_1 R_1 \, , \qquad (52)$$

we find for the phase of the characteristic function (51) from equation (39)

$$\beta_1 \nu_1 + \beta_1^* \nu_1^* = (\alpha \delta + \alpha^* \delta^*) \quad \text{with} \quad \delta = (\nu_1 + \nu_1^*) r_1 R_1 = \delta^* \, . \tag{53}$$

For the variance $\varphi_1^2$, we obtain from equation (40)

$$\varphi_1^2 = 2|\alpha^* + \alpha|^2 r_1 R_1 + |\alpha|^2 (r_2^2 - r_1^2) \tag{54}$$

The characteristic function is that of a canonical ensemble of coherent, squeezed number states with coherent displacement parameter $\delta$ (see Appendix A.1). The parameter $\delta$ is maximal for real $\nu_1$, and vanishes for imaginary $\nu_1$. On the other hand, the variance $\varphi_1^2$ is maximal for real $\alpha$, and minimal for imaginary $\alpha$. The amplified state therefore shows phase squeezing. The amplification and squeezing transformations in the complex $\nu$ plane are illustrated in figure 3.

## 3 Solutions of the model equations

### 3.1 General case

The matrix coefficients $T_{i1}(t)$ are obtained as the classical solutions to equations (1), (2), and (6) with $T_{i1}(0) = \delta_{i1}$. They take the form:

$$T_{i1}(t) = \sum_k \tau_{ik} e^{\lambda_k t} \qquad (i = 1, 2, 3) \tag{55}$$

where $\lambda_k$ are the eigenvalues of the matrix representing the right-hand side of equations (1), (2), and (6) (see appendix A.2). The general form of the eigenvalues depends on the relative size of the gain coefficients $\kappa_i$ ($i = 1, 2$). These coefficients contain a factor involving the frequency of the mode considered. Except for a situation where the Stokes coefficient shows large resonant enhancement over the anti-Stokes coefficient, one has $\kappa_2 > \kappa_1$. We restrict our discussion to this case.

In this case, the eigenvalues are either purely imaginary, or there is an eigenvalue with positive real part. In the former case, the matrix coefficients $T_{i1}$ show periodic or quasi-periodic behavior in time. In the latter case, the term involving an eigenvalue with positive real part dominates (55) for large times, and equation (45) holds. We now discuss some special cases.

## 3.2 Hypertransient limit

In the hypertransient limit, we set $\Gamma = 0$. There are two regimes, determined by a certain critical value $\Delta_{crit}$ of $\Delta$. For $\Delta < \Delta_{crit}$, the anti-Stokes mode dominates. The eigenvalues are all purely imaginary and the system shows quasi-periodic behavior. For $\Delta = 0$, one of the eigenvalues is zero, and the matrix coefficients $T_{i1}(t)$ are periodic, with frequency

$$\kappa = \sqrt{\kappa_2^2 - \kappa_1^2} \tag{56}$$

At times $t$ such that $\cos \kappa t = -1$, we have

$$|T_{11}| = R_1 = \kappa^{-2}(\kappa_1^2 + \kappa_2^2) , \quad |T_{21}| = R_2 = 2\kappa^{-2}\kappa_1\kappa_2 , \quad R_1^2 - R_2^2 = 1$$

This is the special situation mentioned in section 2.2 above. In this case, minimal variance is found for conventional two-mode squeezing $|z_1| = |z_2| = \frac{1}{\sqrt{2}}$:

$$\varphi_{conv}^2 = \varphi_{min}^2 = \frac{1}{2}\left(\frac{\kappa_2 - \kappa_1}{\kappa_2 + \kappa_1}\right)^2 \tag{57}$$

In the following numerical examples we choose $\kappa_1 = 1$ and $\kappa_2 = \sqrt{2}$. Figure (4.1) shows $\varphi_{min}^2$ and $\varphi_{conv}^2$ as functions of $t$ for $\Delta = 0$.

Figure (4.2) illustrates the quasi-periodic behavior of $\varphi_{min}^2$ and $\varphi_{fix}^2$ for $\Delta = 0.290$, just below the critical value $\Delta_{crit} = 0.300$. In this case, two of the eigenvalues are very close to each other, $(\lambda_3 - \lambda_2) = i\epsilon$, $|\epsilon| \ll |\lambda_k|$, and the matrix coefficients $T_{i1}(t)$ are mostly dominated by the corresponding terms with small frequency denominator $\epsilon$. Consequently, the variance $\varphi_{min}^2$ is almost constant, except for short time periods, during which these terms are small due to their explicit time-dependence on $\sin \epsilon t$. The behavior of $\varphi_{fix}^2$ is quite erratic in this case. The variance $\varphi_{conv}^2$ is much greater than $\frac{1}{2}$, except for the short time periods mentioned, and is not shown in this figure.

For $\Delta = \Delta_{crit}$, the matrix coefficients grow linearly in time, while for $\Delta > \Delta_{crit}$, one of the eigenvalues has a positive real part. As mentioned above, the corresponding term in equation (55) dominates the behavior for large times, and the variance $\varphi_{min}^2$ approaches a constant value. For this situation, we also plot the variance $\varphi_{fix}^2$ obtained for a fixed choice of mode coefficients $z_i$, which gives almost minimal variance for large t. Figure (4.3) shows $\varphi^2_{min}$, $\varphi^2_{fix}$ and $\varphi^2_{conv}$ for $\Delta = 0.4$, just above the critical value, while figure (4.4) shows the same for $\Delta = 10$, much larger than $\Delta_{crit}$. In the former case, $\varphi^2_{conv}$ is close to $\varphi^2_{min}$ and less than $\frac{1}{2}$ up to $t = 4$, while in the latter case $\varphi^2_{conv} > \frac{1}{2}$ at all times. In both cases, $\varphi^2_{min}$ and $\varphi^2_{fix}$ go to different values (less than $\frac{1}{2}$) as $t \to \infty$.

## 3.3 Steady-state case

In the steady-state limit, it is assumed that the homogeneous line width $\Gamma$ is much larger than any of the frequencies $\kappa_1$, $\kappa_2$ and $\Delta$. In this limit, the mode operator $b_3$ is approximated by its steady-state value, and we obtain the following equations:

$$\begin{aligned} b_3 &= \Gamma^{-1}(\kappa_1 b_1^\dagger + \kappa_2 b_2 + F) \\ \dot{b}_1^\dagger &= g_{11} b_1^\dagger + g_{12} b_2 + G_1 \\ \dot{b}_2 &= -g_{12} b_1^\dagger - g_{22} b_2 + i\Delta b_2 + G_2 \end{aligned} \qquad (58)$$

where

$$G_i = \kappa_i \Gamma^{-1} F \qquad \text{and} \qquad g_{ij} = \Gamma^{-1} \kappa_i \kappa_j$$

These equations are equivalent to the model studied in references [14] and [16]. The matrix coefficients $T_{i1}$ are given by a formula analogous to (55), but involving only the two eigenvalues corresponding to equation (58). The eigenvalues are:

$$\lambda = -\frac{1}{2}(g_{22} - g_{11} - i\Delta) \pm \sqrt{\frac{1}{4}(g_{22} - g_{11})^2 - \frac{1}{4}\Delta^2 - \frac{1}{2}i\Delta(g_{22} + g_{11})} \quad (59)$$

For $\Delta = 0$, one has:

$$\lambda_1 = 0, \qquad \lambda_2 = -(g_{22} - g_{11}) \qquad (60)$$

and the matrix coefficients have finite limits for $t \to \infty$. In this case, one finds:

$$\begin{aligned} R_1 &\to \kappa^{-2}\kappa_2^2, \qquad R_2 \to \kappa^{-2}\kappa_1\kappa_2 \\ \varphi^2_{fix}, \varphi^2_{min} &\to \frac{\kappa^{-4}}{2}\left(\kappa_2^2(\kappa_1^2 + \kappa_2^2) - \sqrt{\kappa^8 - 2\kappa^6\kappa_2^2 + \kappa^4\kappa_2^4 + 4\kappa_1^2\kappa_2^6}\right) \\ \varphi^2_{conv} &\to \frac{1}{2}\left(\frac{\kappa_2}{\kappa_1 + \kappa_2}\right)^2 < \frac{1}{2} \end{aligned} \qquad (61)$$

Figure (5.1) illustrates this case.

For $\Delta > 0$, one of the eigenvalues is positive, while the other is negative. In this case, the matrix coefficients diverge for $t \to \infty$, and $\varphi_{min}^2$ goes to the finite limit given by equation (47). Figures (5.2) and (5.3) show the cases $\Delta = 0.1$ and $\Delta = 1$. They demonstrate that $\varphi_{min}^2$ and $\varphi_{fix}^2$ converge to the same limit. The conventional variance $\varphi_{conv}^2$ exceeds the vacuum value $\frac{1}{2}$ for sufficiently large time. For $\Delta \gg g_{ik}$, the eigenvalues are given by:

$$\lambda_1 = g_{11} \quad \text{and} \quad \lambda_2 = -g_{22} + i\Delta \tag{62}$$

and the Stokes and anti-Stokes decouple in this limit [14].

## 3.4 Transient case

In the transient case, where $\Gamma$ is of the same order of magnitude as $\kappa_1$ and $\kappa_2$, the cases $\Delta = 0$ and $\Delta \gg \Gamma$ are easily solved. For $\Delta = 0$, the eigenvalues are:

$$\lambda = 0, \qquad \lambda = -\frac{\Gamma}{2} \pm \sqrt{\frac{\Gamma^2}{4} - \kappa^2} \tag{63}$$

and the matrix coefficients converge to the same limits as in the steady-state case. For small $\Gamma$, two eigenvalues have imaginary parts, and additional oscillations occur for small times. For $\Delta \gg \Gamma$, the eigenvalues are, to leading order:

$$\lambda = i\Delta, \qquad \lambda = -\frac{\Gamma}{2} \pm \sqrt{\frac{\Gamma^2}{4} + \kappa_1^2} \tag{64}$$

One eigenvalue has positive real part, and the mode coefficients grow exponentially for large times. For $\Gamma \gg \kappa_1^2$, the results of the steady-state limit are recovered. For intermediate values of $\Delta$, one eigenvalue has positive real part, and the behavior of the variance for large times is similar to the steady-state case.

# 4 Summary and Discussion

For the model of a linear cavity Raman amplifier studied here, we found two main results. First, there exist uncorrelated normal modes, which are obtained from the Stokes and anti-Stokes modes by an anti-squeezing transformation. Second, there exist optimal linear combination modes, for which the variance of one quadrature is always below the vacuum value. The states of such optimal modes are canonical ensembles of coherent squeezed number states, and the average number of noise photons is proportional to the square root of the number of Stokes noise photons. Thus, the degree of decoherence for the combination mode is much less than for the individual Stokes and anti-Stokes modes.

The reduction of variance in the minimal quadrature of the optimal combination mode can be understood as resulting from a cancellation of anti-correlated quantum noise in the individual modes. To optimize the amount of cancellation, the mode coefficients have to be chosen appropriately, and will in general not be of equal magnitude. This corresponds to a generalization of the conventional notion of two-mode squeezing.

If a coherent state is used as input to the Stokes channel of the amplifier, the corresponding field amplitude will be cancelled in the combination mode, if it has the same phase as the minimal quadrature, and will be optimally amplified if its phase differs by $\pm 90$ degrees. The optimally amplified signal is thus phase-squeezed.

These results may be of interest for the generation of squeezed light through stimulated Raman scattering. They also provide clear experimental signatures for the experimental observation of anti-correlation between Stokes and anti-Stokes. It is of particular interest, that these signatures are found for all values of the experimentally relevant parameters, such as pump amplitude, dispersion and line widths. Previous work on quadrature squeezing in this model had been restricted to a limited range of parameters (small linewidth, zero dispersion).

With the recent demonstration of a cw-Raman laser by J.K. Brasseur et al. [19], experiments to verify our predictions may become feasible in the near future.

# Acknowledgments

It is our pleasure to acknowledge receipt of a preprint "Continuous wave Raman laser in $H_2$" (Ref [19]) from J.L. Carlsten.

# Appendix A.1

The quantum-mechanical density matrix $\rho$ of a system with creation and annihilation operators $a^\dagger$ and $a$ is uniquely characterized by its Wigner characteristic function

$$\chi(\alpha, \alpha^*) = \text{Trace}\left(\rho\, e^{i\phi(\alpha,\alpha^*)}\right) \equiv\, <e^{i\phi(\alpha,\alpha^*)}> \tag{65}$$

where

$$\phi(\alpha, \alpha^*) = \alpha a + \alpha^* a^\dagger \ .$$

Of special interest are Gaussian states, for which

$$\chi = e^{i<\phi>-\frac{1}{2}<(\Delta\phi)^2>}\ ,\quad \Delta\phi = \phi - <\phi>\ . \tag{66}$$

The most general Gaussian state is a canonical ensemble of coherent squeezed number states. Using the notation of Caves [1], we write this as

$$\rho = D(\delta)T(\zeta)\rho_{th}T^{-1}(\zeta)D^{-1}(\delta)$$

$$\text{with}\quad D(\delta) = e^{\delta a - \delta^* a^\dagger}\ ,\quad T(\zeta) = e^{\frac{1}{2}\zeta^* a^2 - \frac{1}{2}\zeta a^{\dagger 2}}\ ,\quad \zeta = se^{i\theta} \tag{67}$$

$$\text{and}\quad \rho_{th} = (1 - e^{-r})\sum_n e^{-rn}|n><n|\ .$$

For the expectation value and variance of $\phi$, one finds [1]:

$$<\phi> = \alpha\delta + \alpha^*\delta^*$$
$$<(\Delta\phi)^2> = Q^2|\alpha U + \alpha^* V|^2 \tag{68}$$

with:

$$Q^2 = 2n + 1 = \frac{1 + e^{-r}}{1 - e^{-r}}\ ;\quad n = <a^\dagger a> = \frac{e^{-r}}{1 - e^{-r}}$$

$$U = \cosh s\ ,\quad V = -e^{i\theta}\sinh s \tag{69}$$

Equations (66), (68) and (69) give the characteristic function in terms of the coherent-state displacement parameter $\delta$, the squeezing parameter $\zeta$ through U and V and the mean photon number n through Q as:

$$\chi(\alpha, \alpha^*) = e^{i(\alpha\delta + \alpha^*\delta^*) - \frac{1}{2}Q^2|\alpha U + \alpha^* V|^2}\ . \tag{70}$$

## Appendix A.2

The matrix coefficients $T_{i1}(t)$ for $i = 1, 2, 3$ satisfy equations (1),(2) and (6), with initial conditions

$$T_{11}(0) = 1 \quad \text{and} \quad T_{i1}(0) = 0 \quad \text{for} \quad i = 2, 3 \tag{71}$$

They take the form

$$T_{i1}(t) = \sum_k \tau_{ik} e^{\lambda_k t} \quad i = 1, 2, 3 \tag{72}$$

where $\lambda_k$ are the eigenvalues of the matrix of coefficients $N$ for equations (1),(2) and (6):

$$N = \begin{pmatrix} 0 & 0 & \kappa_1 \\ 0 & i\Delta & -\kappa_2 \\ \kappa_1 & \kappa_2 & -\Gamma \end{pmatrix} \tag{73}$$

We find:

$$\tau_{1k} = \frac{(\lambda_i \lambda_j + \kappa_1^2)}{d_k} \qquad \tau_{2k} = -\frac{\kappa_1 \kappa_2}{d_k} \qquad \tau_{3k} = \frac{\kappa_1^2 (\Gamma - S_k)}{d_k} \tag{74}$$

where

$$d_k = (\lambda_k - \lambda_i)(\lambda_k - \lambda_j) \quad \text{and} \quad S_k = \lambda_i + \lambda_j$$

and, for fixed k, the indices $i$ and $j$ take the two values different from k. The eigenvalues were obtained numerically.

Analytical solutions or approximations for the eigenvalues were obtained in the special cases considered in subsections (3.2) and (3.3), in particular the cases: $\Gamma = 0, \ \Gamma \gg \Delta, \kappa_i^2, \ \Delta = 0, \ \Delta \gg \Gamma, \kappa_i^2,$ and were used to verify the correctness of the numerical procedure. The calculations are straightforward and will not be detailed here.

# Figure Captions

Figure 1. Shown are the two levels of the Raman transition at frequencies $\omega = 0$ and $\omega = \omega_3$, and two intermediate levels at frequencies $\omega = \omega_0$ and $\omega = \omega_0 + \omega_3$ for Stokes and anti-Stokes generation. The corresponding absorption of pump radiation and generation of Stokes and anti-Stokes is indicated by vertical and horizontal arrows. The anti-Stokes mode at frequency $\omega = \omega_0 + \omega_3 + \Delta$ and its driving polarization at $\omega = \omega_0 + \omega_3$ are represented by two different lines, separated by the frequency detuning $\Delta$.

Figure 2 illustrates the sequence of entangling and dis-entangling operations performed by the cavity Raman laser (CRL) and linear parametric amplifier (LPA). The uncorrelated states of the initial Stokes and anti-Stokes modes $a_1$ and $a_2$, and of the final normal modes $c_1$ and $c_2$, are represented by disjoint circles, while the entangled state of the intermediate, amplified Stokes and anti-Stokes modes $b_1$ and $b_2$ is represented by a single ellipse. The quantities $n_1$ and $n_2$ are the corresponding photon numbers.

Figure 3 illustrates the effect of amplification on the coherent states of the near-optimal combination mode $b$. The coherent states are represented by ellipses in the complex $\nu$ plane, which are centered at the coherent displacement parameter $\nu = \nu_1$ for the initial states and $\nu = \delta$ for the final states. The axes of the ellipses are proportional to the root of the variances of the corresponding quadratures. The arrows connect the initial state to the final state resulting from amplification. Case (a) represents the vacuum going to a squeezed state. Cases (b), (c) and (d) represent various initial coherent states with imaginary, real and complex $\delta$ going the corresponding final states, with (b) going to the same state as (a), and (c) going to the same state as (d).

Figures 4.a to 4.d. Shown are the variances of the minimal quadrature variable for the choice of conventional, fixed and minimizing mode coefficients ($\varphi^2_{\text{conv}}$, $\varphi^2_{\text{fix}}$ and $\varphi^2_{\text{min}}$) at zero linewidth ($\Gamma = 0$), for various amounts of dispersive detuning as given in the figures ($\Delta = 0.0$, $\Delta = 0.29$, $\Delta = 0.4$ and $\Delta = 10.0$).

Figures 5.a to 5.c. Shown are the variances of the minimal quadrature variable for the choice of conventional, fixed and minimizing mode coefficients

($\varphi^2_{\text{conv}}$, $\varphi^2_{\text{fix}}$ and $\varphi^2_{\text{min}}$) at large linewidth ($\Gamma = 5$), for various amounts of dispersive detuning as given in the figures ($\Delta = 0.0$, $\Delta = 0.1$ and $\Delta = 1.0$).

# 5  Figures

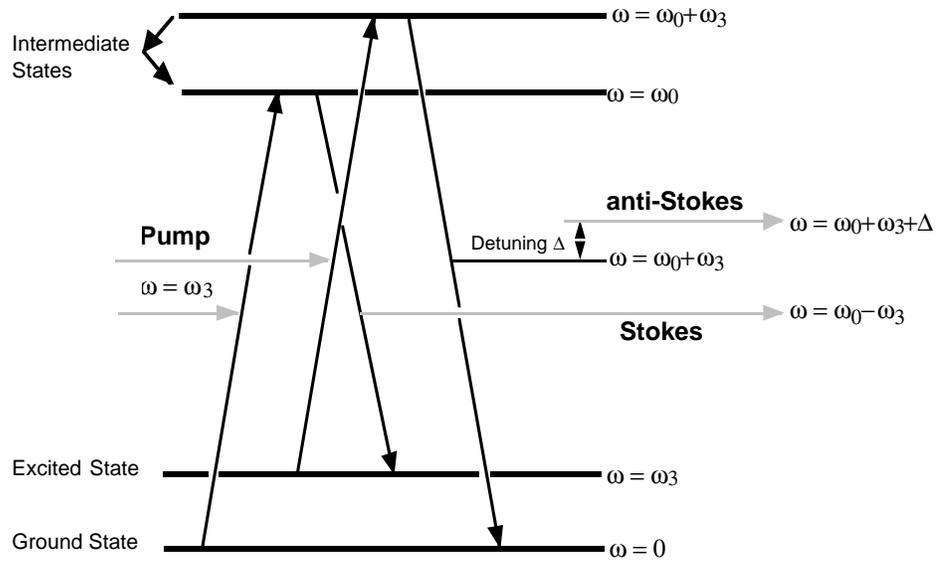

Figure 1

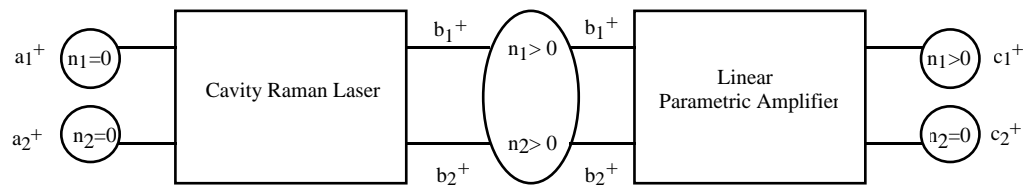

Figure 2

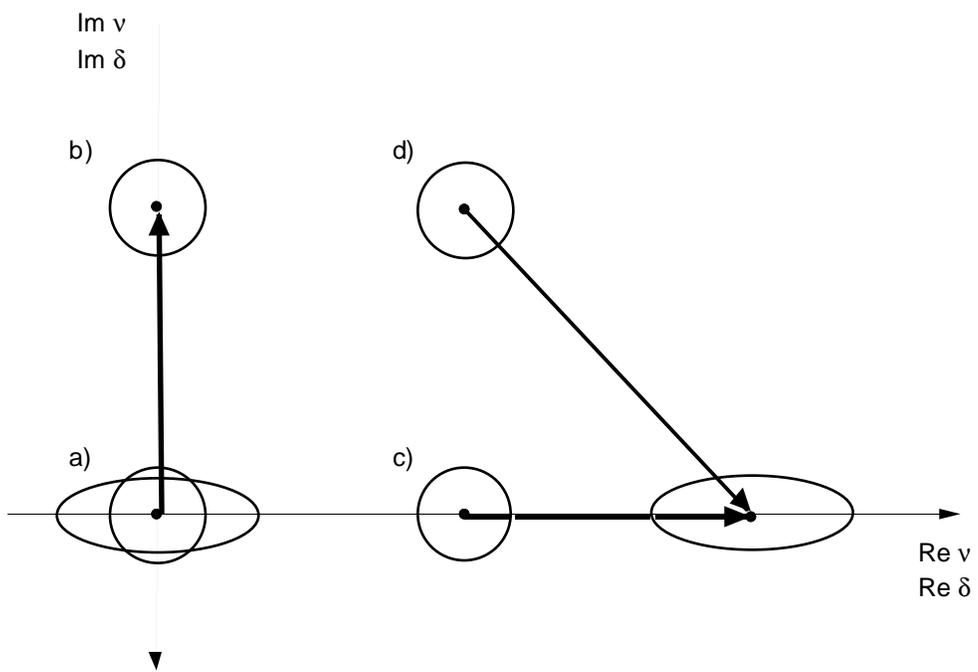

Figure 3

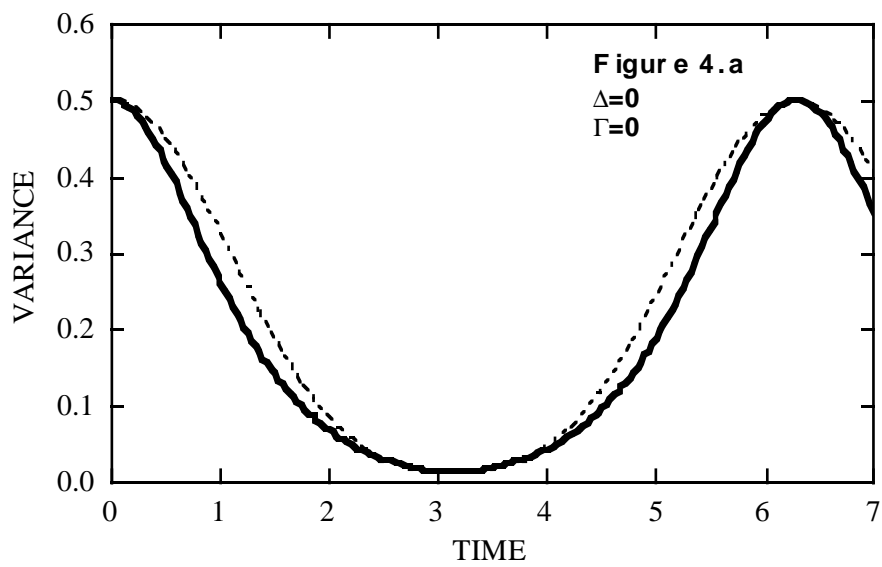

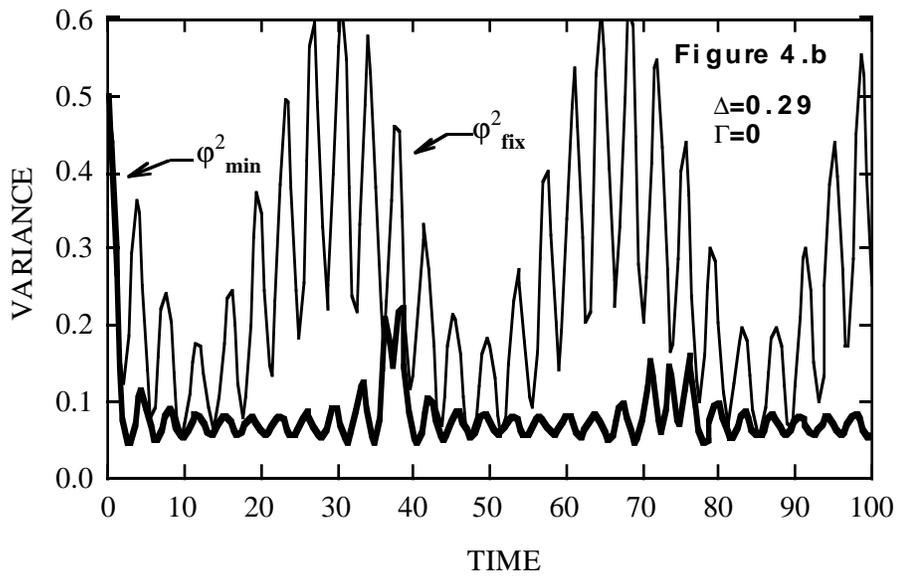

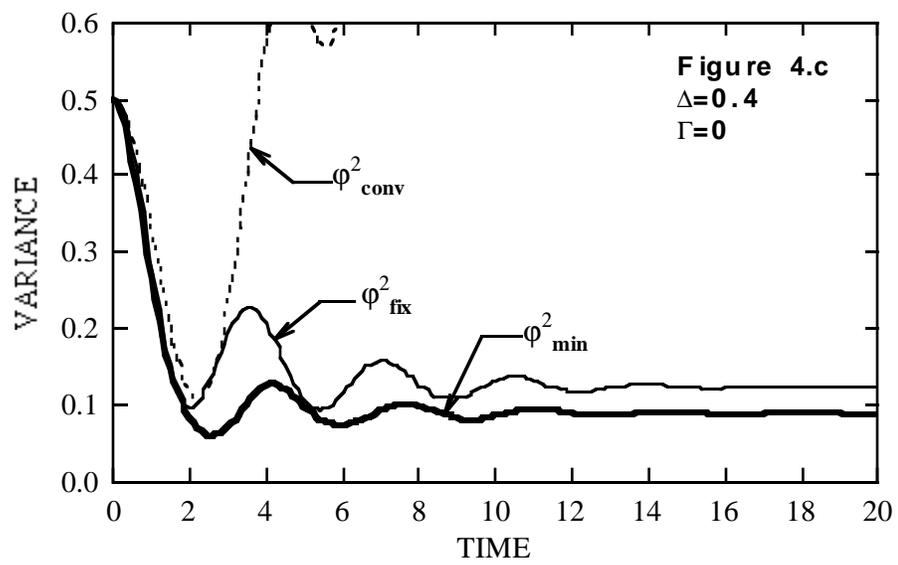

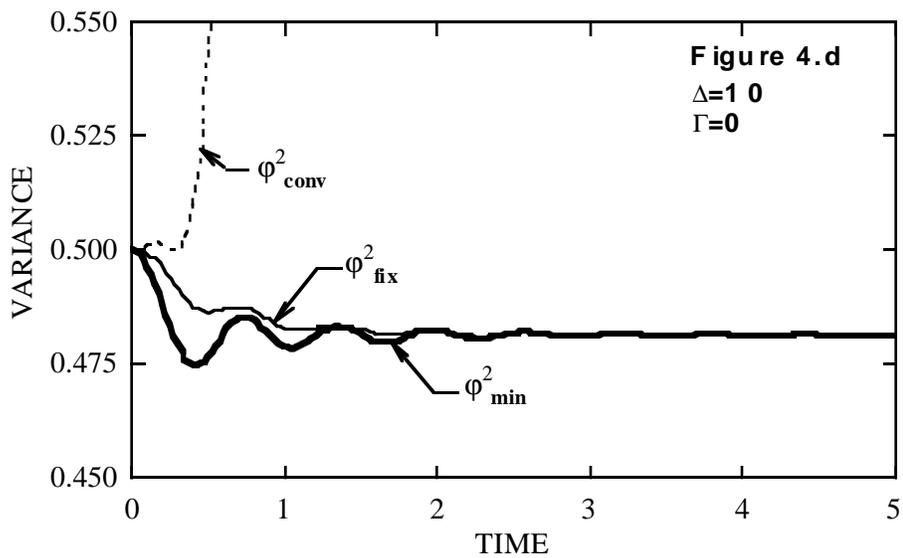

Figure 4.d
Δ=10
Γ=0

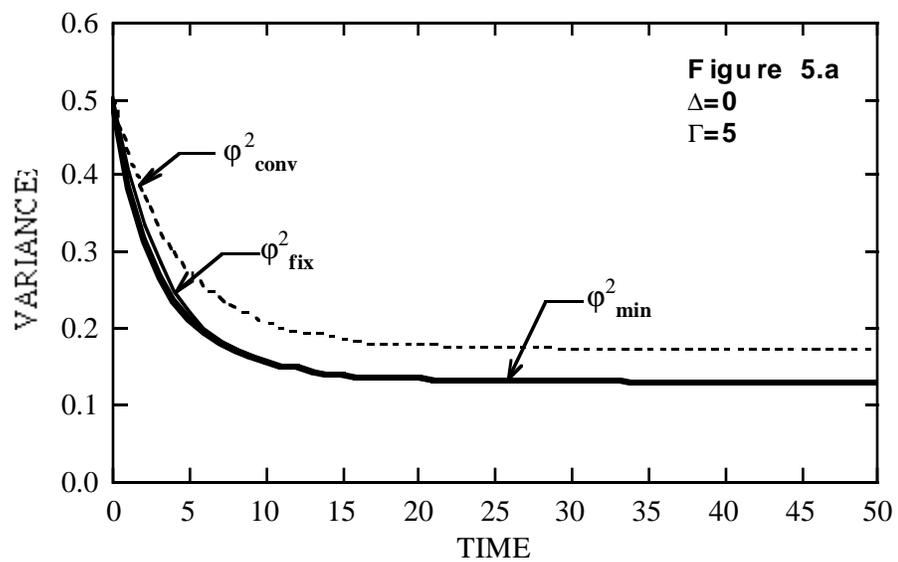

Figure 5.a
Δ=0
Γ=5

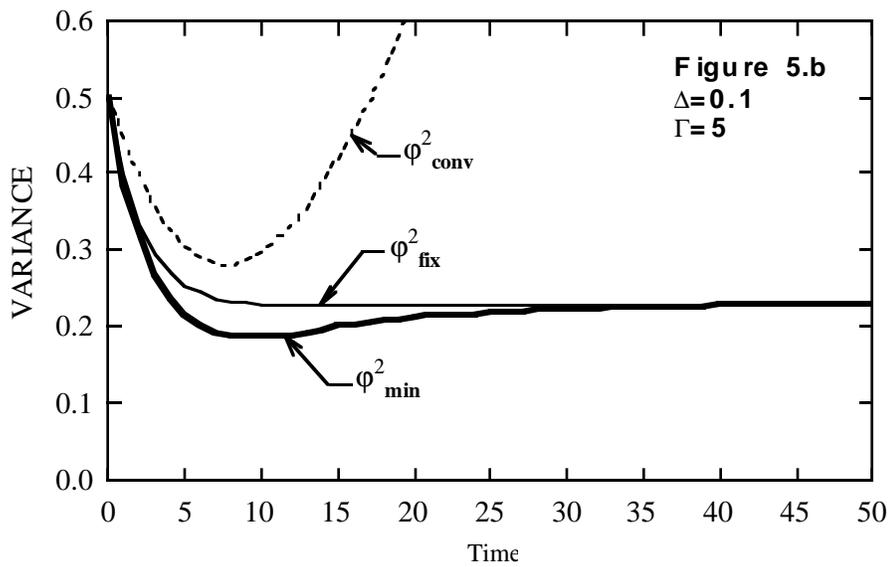

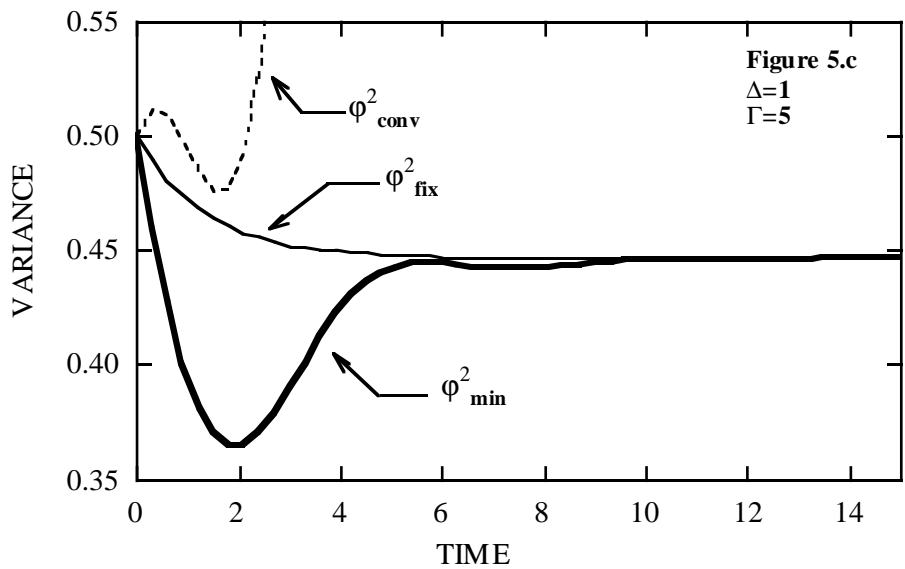